# Revealing the Origin of Luminescence Center in 0D Cs$_4$PbBr$_6$ Perovskite

Zhaojun Qin[1,2,†], Shenyu Dai[2,3,†], Viktor G. Hadjiev[4,5,*], Chong Wang[6,2], Lingxi Ouyang[6,2], Lixin Xie[4,7], Yizhou Ni[4,7], Chunzheng Wu[1,2], Guang Yang[8], Shuo Chen[4,7], Liangzi Deng[4,7], Qingkai Yu[9], Ching-Wu Chu[4,7,10], Guoying Feng[3,*], Zhiming Wang[1,*], Jiming Bao[2,8,*]

[1]Institute of Fundamental and Frontier Sciences, University of Electronic Science and Technology of China, Chengdu, Sichuan 610054, China

[2]Department of Electrical and Computer Engineering, University of Houston, Houston, Texas 77204, USA

[3]College of Electronics & Information Engineering, Sichuan University, Chengdu, Sichuan 610064, China

[4]Texas Center for Superconductivity, University of Houston, Houston, TX 77204, USA

[5]Department of Mechanical Engineering, University of Houston, Houston, TX 77204, USA

[6]School of Materials Science and Engineering, Yunnan University, Kunming, Yunnan 650091, China

[7]Department of Physics, University of Houston, Houston, Texas 77204, USA

[8]Materials Science & Engineering, University of Houston, Houston, Texas 77204, USA

[9]Ingram School of Engineering, Texas State University, San Marcos, Texas 78666, USA

[10]Lawrence Berkeley National Laboratory, Berkeley, California 94720, USA

†Equal contributions

*To whom correspondence should be addressed: Jiming Bao (jbao@uh.edu),

Zhiming Wang (zhmwang@uestc.edu.cn), Viktor Hadjiev (vhadjiev@central.uh.edu), Guoying Feng (guoing_feng@scu.edu.cn).

# Abstract

Zero dimensional perovskite $Cs_4PbBr_6$ has attracted considerable attention recently not only because of its highly efficient green photoluminescence (PL), but also its two highly debated opposing mechanisms of the luminescence: embedded $CsPbBr_3$ nanocrystals versus intrinsic Br vacancy states. After a brief discussion on the root cause of the controversy, we provide sensitive but non-invasive methods that can not only directly correlate luminescence with the underlying structure, but also distinguish point defects from embedded nanostructures. We first synthesized both emissive and non-emissive $Cs_4PbBr_6$ crystals, obtained the complete Raman spectrum of $Cs_4PbBr_6$ and assigned all Raman bands based on density functional theory simulations. We then used correlated Raman-PL as a passive structure-property method to identify the difference between emissive and non-emissive $Cs_4PbBr_6$ crystals and revealed the existence of $CsPbBr_3$ nanocrystals in emissive $Cs_4PbBr_6$. We finally employed a diamond anvil cell to probe the response of luminescence centers to hydrostatic pressure. The observations of fast red-shifting, diminishing and eventual disappearance of both green emission and Raman below $Cs_4PbBr_6$ phase transition pressure of ~3 GPa is compatible with $CsPbBr_3$ nanocrystal inclusions as green PL emitters and cannot be explained by Br vacancies. The resolution of this long-lasting controversy paves the way for further device applications of low dimensional perovskites, and our comprehensive optical technique integrating structure-property with dynamic pressure response is generic and can be applied to other emerging optical materials to understand the nature of their luminescent centers.

The lack of deep-level, carrier trapping and defect states in $CsPbX_3$ (X=I, Br, Cl) determines in part the defect-tolerant electronic and optical properties of these all-inorganic perovskites and make them auspicious materials for high-efficiency low-cost solar cells and many other optoelectronic devices[1, 2, 3, 4]. As such, recent observations of apparently deep-level and highly luminescent states in low-dimensional lead halide perovskites such as 0D $Cs_4PbBr_6$[5, 6, 7, 8, 9] and 2D $CsPb_2Br_5$[10, 11, 12, 13, 14, 15], have attracted a lot of attention as well as intensive debates. Among them, the debate on the origin of bright green luminescence in the otherwise wide bandgap $Cs_4PbBr_6$ is more intense and involving a large community, as can be seen from four recent critical reviews representing two opposing opinions[6, 7, 8, 9]. Because of the extreme similarity of its green emission with that of $CsPbBr_3$ nanocrystals, it is believed that the embedded $CsPbBr_3$ nanocrystals are responsible for the highly efficient photoluminescence (PL) [8, 9, 16, 17, 18, 19]. This opinion is further strengthened by direct imaging of $CsPbBr_3$ nanocrystals in $Cs_4PbBr_6$ [20, 21, 22, 23, 24, 25, 26, 27, 28, 29]. Nevertheless, the other group offers a totally different theory, attributing the strong PL to Br vacancies and regarding the green emission as an intrinsic property of $Cs_4PbBr_6$ [5, 7, 30, 31, 32, 33, 34, 35, 36]. This Br vacancy theory is supported by their density functional theory (DFT) simulation and most importantly, the observation of pure $Cs_4PbBr_6$ single crystals without embedded $CsPbBr_3$ nanocrystals by high resolution transmission electron microscopy (TEM)[7]. It is also supported by their observation that the appearance of $CsPbBr_3$ in initially bright $Cs_4PbBr_6$ will quench its PL[30], although this counter-intuitive observation has been questioned[8]. The Br vacancy theory also has problem because the calculated deep-level defect states are not reproduced by other DFT simulation and are rarely found experimentally in lead halide perovskites[9].

Structure-property relation is the essential goal of materials science. The unsettling of this controversy indicates the challenge of controversy and limitation of existing efforts. Although simulation has become more powerful in predicting material properties, experimental evidence is still pivotal to fundamental understanding and is the test for any theoretical contenders. The root cause for the dispute is the lack of one-to-one correlation between luminescence and structure, and an experimental technique that can distinguish luminescence from point defects versus nano-inclusions. For example, the structure of a single $Cs_4PbBr_6$ nanocrystal was confirmed by high resolution TEM, but the PL of the same exact nanoparticle was not reported or confirmed, even though this nanocrystal was selected from an ensemble of emissive nanocrystals[7, 14, 37, 38], so one-to-one relationship was not solidly established[15]. On the other hand, it is well known that lead halide perovskites are very sensitive to electron beams and can easily get damaged[39], thus the TEM evidence from both sides could be questionable. Here we report a resolution of this controversy and identification of the origin of luminescent centers in $Cs_4PbBr_6$. This has been achieved by using a combined confocal Raman-PL technique in conjunction with a diamond anvil cell (DAC) that can directly correlate structural information with luminescent property at the same length scale. Since the complete Raman spectrum of $Cs_4PbBr_6$ was unknown at the time of this study, we recorded and assigned all Raman active modes. The response of PL and Raman scattering of $Cs_4PbBr_6$ to hydrostatic pressure helped us to elucidate whether the PL comes from point defect or extended structures.

We synthesized both green PL emissive and non-emissive $Cs_4PbBr_6$ (shortly emissive and non-emissive) based on reported methods so that similar experiments from opposing sides can be checked under the same conditions[28, 30, 31]. $CsPbBr_3$ micro-powders and highly luminescent $CsPbBr_3$ nanocrystals were also prepared as PL and Raman references for comparison[16, 40, 41]. Figs. 1a-b show optical images of emissive and non-emissive $Cs_4PbBr_6$ nanocrystal suspensions. The non-emissive $Cs_4PbBr_6$ nanocrystals appear transparent, confirming it as a wide bandgap semiconductor. However, no difference in their X-ray (XRD) patterns (Fig. SF1) is observed[5, 7, 30, 31, 32, 33, 34, 35, 36] except a wider linewidth of non-emissive $Cs_4PbBr_6$ due to its smaller crystal size (Fig. SF2). Like emissive $Cs_4PbBr_6$ nanocrystals, large size $Cs_4PbBr_6$ crystals also exhibit a yellow color under white light and strong green color under UV. To overcome the low sensitivity of XRD, we employed TEM to identify possible $CsPbBr_3$ nano-inclusions in $Cs_4PbBr_6$. In contrast to previous work[7, 14, 37, 38], here we performed TEM and PL on the same emissive single crystals[42]. Figs. 1e-l confirm their excellent $Cs_4PbBr_6$ structure with clear crystal facets, and similar to XRD, no embedded $CsPbBr_3$ nanocrystals were detected. However, it is important to point out that after exposure to electron beams, green emission from these nanocrystals either decreased significantly or disappeared, indicating a severe structural damage.

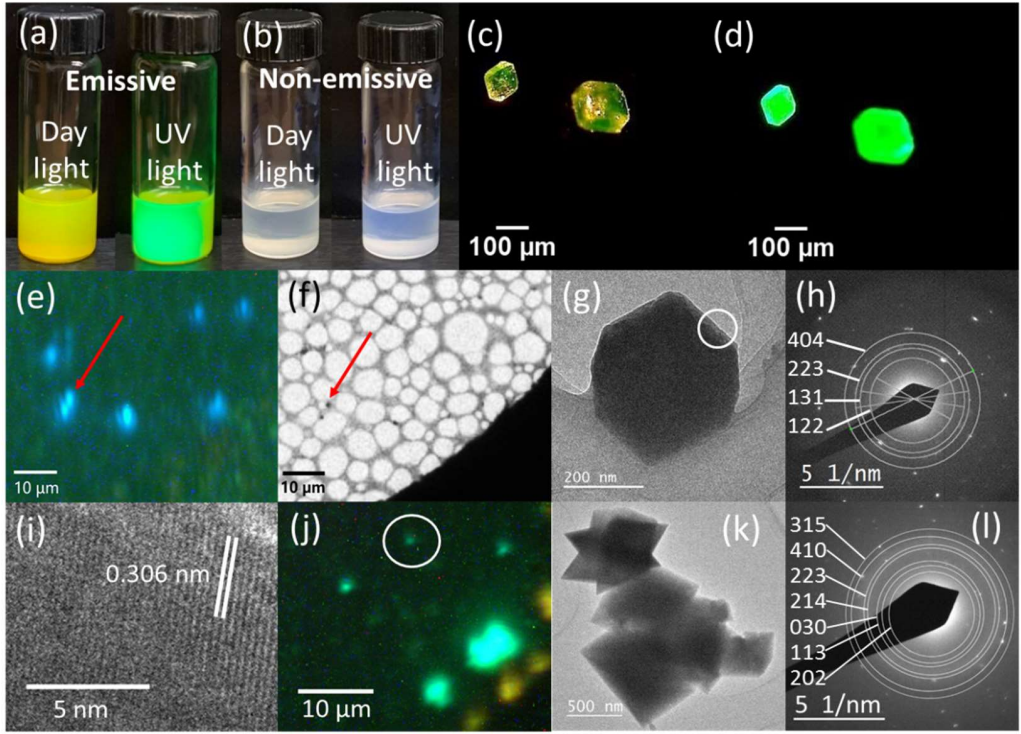

Fig. 1. Failure to detect $CsPbBr_3$ nanocrystals in $Cs_4PbBr_6$ with transmission electron microscopy (TEM). (a-b) Optical images of emissive and non-emissive $Cs_4PbBr_6$ nanocrystals. (c-d) Optical images of large size emissive $Cs_4PbBr_6$ single crystals under (c) white light and (d) UV light. (e-i) Photoluminescence (PL) image, low and high magnification TEM images, selected area (indicated by the white circle in (g) electron diffraction and lattice fringe images of an emissive $Cs_4PbBr_6$ nanocrystal (indicated by the red arrows in (e) and (f)). (j-l) PL, TEM and electron diffraction images of $Cs_4PbBr_6$ nanocrystals obtained from large single crystals as in (c-d).

After failing to detect $CsPbBr_3$ nanocrystals in $Cs_4PbBr_6$, we turned to Raman spectroscopy, a non-invasive and sensitive optical technique. Since to the best of our knowledge there has been no report on the complete Raman spectra of $Cs_4PbBr_6$ but only partial ones[32, 43], we provide a more detailed analysis of the experimental Raman spectrum. In addition to the need of proper assignment of Raman active phonons, this analysis is helpful in discerning intrinsic from inclusion Raman features. We choose non-emissive $Cs_4PbBr_6$ as a reference for Raman spectrum. Its Raman at 80 K is shown in Fig. 2a.

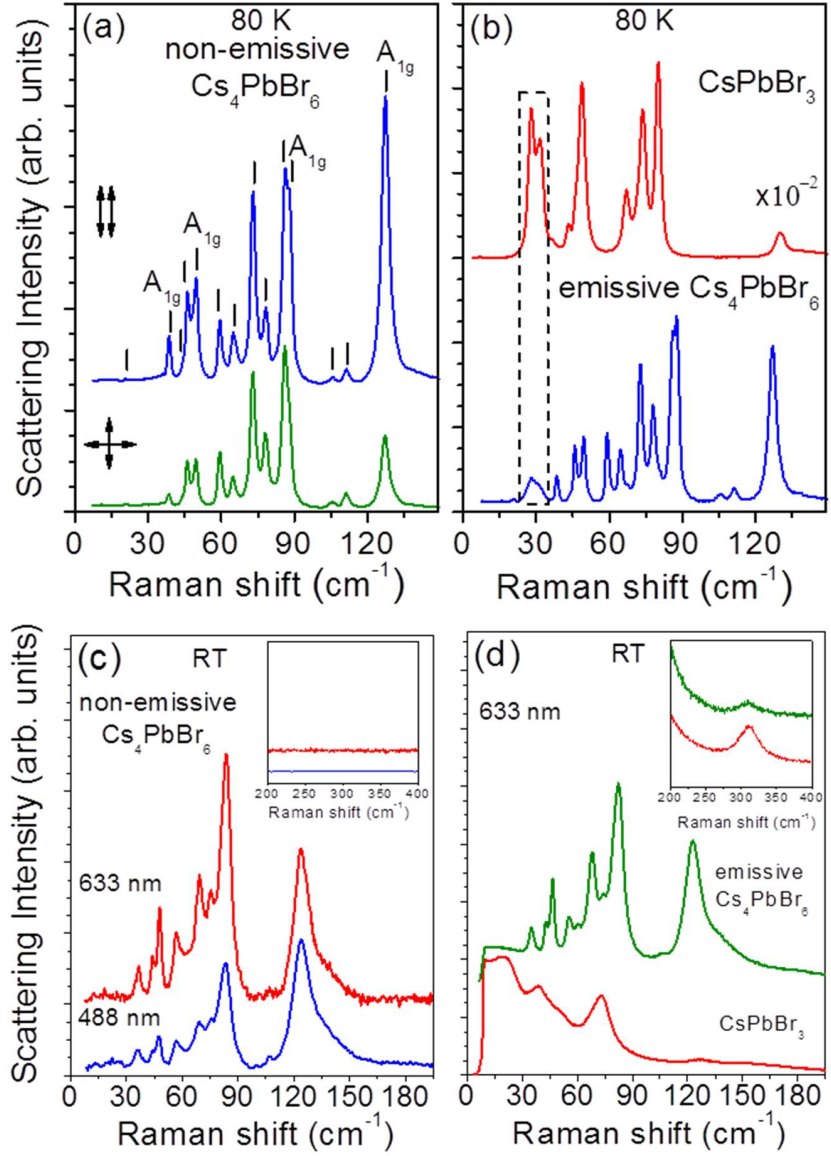

Fig. 2. Raman spectra of CsPbBr$_3$ microcrystals, non-emissive and emissive Cs$_4$PbBr$_6$ at 80 K and room temperature (RT). (a) Polarized spectra of non-emissive Cs$_4$PbBr$_6$ at 80 K with all $4A_{1g} + 10E_g$ Raman modes marked with vertical bars. ∥ stands for parallel and + for orthogonal incident $\vec{e}_I$ and scattered $\vec{e}_S$ light polarizations, respectively. (b) Raman spectra of emissive Cs$_4$PbBr$_6$ and CsPbBr$_3$ at 80 K. (c) Raman spectra of non-emissive Cs$_4$PbBr$_6$ at RT. (d) Raman spectra of emissive Cs$_4$PbBr$_6$ at RT. The Raman spectrum of CsPbBr$_3$ at RT is also included for reference. The presence of CsPbBr$_3$ in emissive Cs$_4$PbBr$_6$ is evidenced by the second order Raman band at 310 cm$^{-1}$ in (d) and Cs phonon peaks of CsPbBr$_3$ at 80 K. Raman spectra in (a-b) and (d) are excited with 632.8-nm laser. The scattering intensity scale is the same for all spectra in a given panel.

Cs$_4$PbBr$_6$ crystalizes in a trigonal crystal structure characterized by the space group $R\bar{3}c$ (No. 167)[44]. The rhombohedral unit cell is the primitive cell (PC) of this compound and contains two formula units of Cs$_4$PbBr$_6$, $N_{cell}$ = 22 atoms per PC and $3N_{cell}$ = 66 degrees of vibrational freedom. The irreducible representations of the Γ-point phonon modes are $4A_{1g} + 10E_g + 6A_{2g} + 5A_{1u} + 7A_{2u} + 12E_u$ and only the $A_{1g}$ and $E_g$ phonons are Raman active[21]. The three acoustic phonons are represented by $A_{2u} + E_u$ symmetry modes, the remaining $5A_{2u} + 11E_u$ modes are infrared (IR) active, whereas $5A_u + 6A_{2g}$ modes are neither IR nor Raman active. One experimental approach to discerning different symmetry modes is to measure the polarized Raman spectra. The Raman tensor, $R_S = |\alpha_{ij}|$ with $i, j = x, y, z$, of active modes $S = A_{1g}$ and $E_g$, has the following non-zero components[45]: $R_{A_{1g}}(\alpha_{xx} = \alpha_{yy} = a, \alpha_{zz} = b)$, $R_{E_{g,1}}(\alpha_{xx} = -\alpha_{yy} = c, \alpha_{yz} = \alpha_{zy} = d)$, and $R_{E_{g,2}}(\alpha_{xy} = \alpha_{yx} = -c, \alpha_{xz} = \alpha_{zx} = -d)$. The analysis of Raman scattering activity $I_S = [\vec{e}_S \cdot R_S \cdot \vec{e}_I]^2$, where $\vec{e}_I$ and $\vec{e}_S$ are the incident and scattered light polarizations, suggests that for polycrystalline or randomly oriented nanocrystals of Cs$_4$PbBr$_6$[46], as in the case of most of our samples, the depolarization ratio for $A_{1g}$ modes is $0 \leq \frac{I^+_{A_{1g}}}{I^\parallel_{A_{1g}}} \leq \frac{3}{4}$ and $\frac{I^+_{A_{1g}}}{I^\parallel_{A_{1g}}} = \frac{3}{4}$ for the $E_g$ modes, where ∥ stands for parallel and + for orthogonal $\vec{e}_I$ and $\vec{e}_S$, respectively. This allowed us to distinguish the four $A_{1g}$ modes in Fig. 2a by their markedly smaller depolarization ratio. Our experimental assignment of the Raman modes is further confirmed by DFT lattice dynamics calculations as seen in Table SF1.

With Raman standard of Cs$_4$PbBr$_6$ at hand, we can identify the Raman difference between emissive and non-emissive Cs$_4$PbBr$_6$. As shown in Fig. 2b, the Raman spectrum of emissive Cs$_4$PbBr$_6$ is identical to that of non-emissive Cs$_4$PbBr$_6$ in Fig. 2a except that the spectrum of emissive Cs$_4$PbBr$_6$ contains an additional Raman band at ~29 cm$^{-1}$ that replicates the doublet at 28-30 cm$^{-1}$ in CsPbBr$_3$[47, 48], indicating presence of CsPbBr$_3$ in Cs$_4$PbBr$_6$. Figure SF9 shows the difference in atomic displacements of Cs that are involved in 28-30 cm$^{-1}$ doublet in CsPbBr$_3$ and those in Cs4PbBr$_6$. Under 632.8 nm laser excitation the Raman scattering in both compounds is non-resonant. This allowed us to estimate trace amount of CsPbBr$_3$ in the emissive Cs$_4$PbBr$_6$. Based on the relative Raman intensity, the concentration of CsPbBr$_3$ is estimated to 0.2 % by volume, which is below typical XRD sensitivity.

The observation of Raman difference and identification of CsPbBr$_3$ in Cs$_4$PbBr$_6$ proves that Raman is the right method to solve this controversy, however, the low temperature used is usually not very convenient; hence we explore room temperature Raman spectra for finer details. Fig. 2c shows the RT Raman spectra of non-emissive Cs$_4$PbBr$_6$ excited with 632.8 nm (1.96.eV) and 488 nm (2.54 eV) laser lines. The spectra are identical to that of low temperature except that lines become broader. The good signal-to-noise ratio and lack of strong background in the spectrum excited with the 488 nm laser demonstrate that no green PL is excited in this sample. Fig. 2d shows Raman spectrum of emissive Cs$_4$PbBr$_6$. The feature at ~29 cm$^{-1}$ stemming from CsPbBr$_3$ becomes indistinguishable, but this agrees with the stronger background due to dynamic-disorder scattering from Cs coupled to Br anharmonic motion in CsPbBr$_3$ at RT[47], as also shown in Fig. 2d. A close look at the 200 – 400 cm$^{-1}$ spectral range displayed in the insets of Figs. 2c and 2d, however, reveals a Raman band around 310 cm$^{-1}$ in the emissive sample, whereas this band is absent in the non-emissive one. The band at 310 cm$^{-1}$ is at the same position as in the second-order phonon Raman scattering in CsPbBr$_3$[32, 47, 48] also shown in Fig. 2d for comparison.

Because of no Raman spectral overlapping with Cs$_4$PbBr$_6$ and relatively large wavenumber, the 310 cm$^{-1}$ line is a convenient spectral signature to identify CsPbBr$_3$ at RT. However, due to the low concentration of CsPbBr$_3$ in Cs$_4$PbBr$_6$ and weak intensity of second-order phonon Raman scattering, the detection of 310 cm$^{-1}$ line requires a high signal to noise ratio. To better understand this challenge, we chose CsPbBr$_3$ nanocrystals (100 nm in average diameter, Fig. SF3) and emissive Cs$_4$PbBr$_6$ and performed confocal Raman-PL on the same spot and under the same condition. Fig. 3a shows PL of CsPbBr$_3$ nanocrystals and Cs$_4$PbBr$_6$ under 473-nm laser excitation, and their same spot Raman spectra with 632.8 nm laser are shown in Fig. 3b. Both samples exhibit very similar PL, with slightly stronger PL for Cs$_4$PbBr$_6$, but the 310 cm$^{-1}$ CsPbBr$_3$ feature in Cs$_4$PbBr$_6$ is orders of magnitude weaker than in the CsPbBr$_3$ nanocrystals as seen in the inset of Fig. 3b. Because of higher PL quantum yield for smaller CsPbBr$_3$, the stronger PL but much weaker Raman indicates that CsPbBr$_3$ in Cs$_4$PbBr$_6$ must be much smaller than the 100-nm CsPbBr$_3$ nanocrystals, in agreement with the TEM of Cs$_4$PbBr$_6$ nanocrystals in Fig. 2.

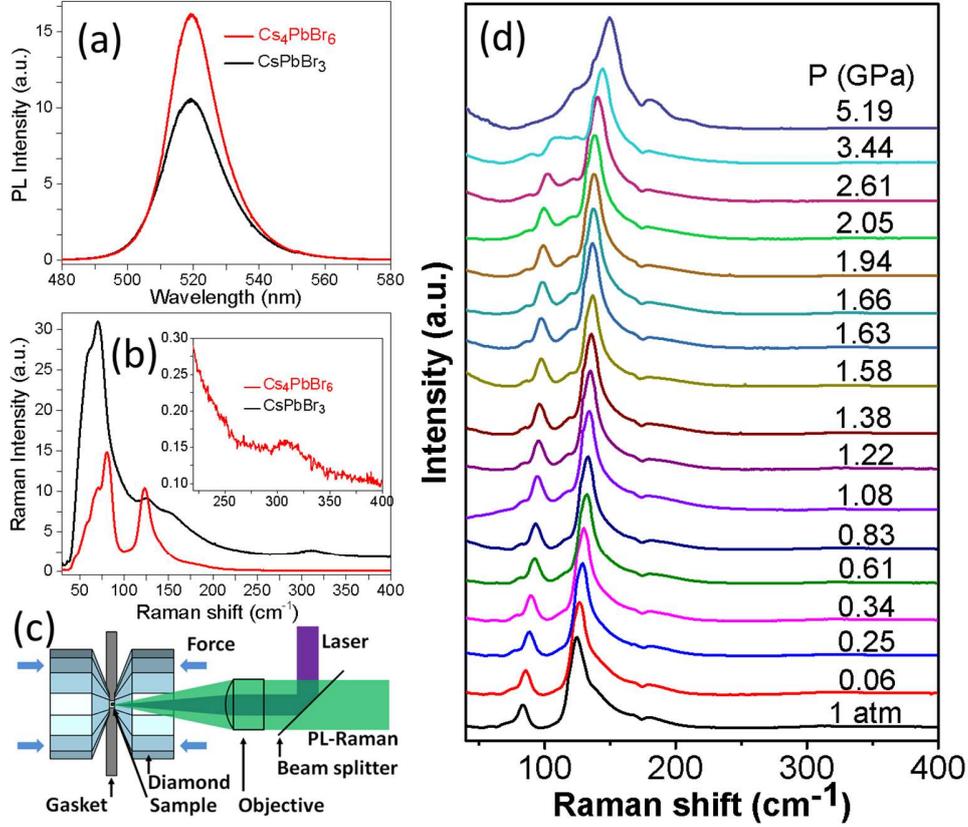

Fig. 3. Room temperature PL and Raman of CsPbBr$_3$ nanocrystals and emissive Cs$_4$PbBr$_6$. (a) PL spectra under 473-nm laser excitation. (b) Raman spectra under 633-nm laser excitation. (c) Diamond anvil cell for confocal pressure Raman-PL. (d) Pressure evolution of Raman of emissive Cs$_4$PbBr$_6$. The weaker appearance of closely spaced $E_g$ and $A_{1g}$ lines at 88 cm$^{-1}$ is due their closeness to the spectral cutoff of a dichroic beam splitter.

The observation of CsPbBr$_3$ Raman in emissive Cs$_4$PbBr$_6$ is a significant step toward the resolution of the controversy, however, it cannot completely rule out the existence of other green emitters such as Br vacancies suggested as a sole source of green PL in Cs$_4$PbBr$_6$[7]. In order to distinguish between presumably different green PL emitters, we study the Raman and PL response of Cs$_4$PbBr$_6$ to hydrostatic pressure using a high-throughput Raman/PL spectrometer. Fig. 3c shows the schematic of experimental setup, and Fig. 3d shows the evolution of Raman under pressure to up to 5 GPa. The Raman is still dominated by two peaks, the weaker appearance of $E_g/A_{1g}$ line at 88 cm$^{-1}$ is due to the spectral cutoff of a dichroic beam splitter, a complete Raman spectrum evolution recorded on a triple spectrometer with cutoff at 10 cm$^{-1}$ is shown in Fig. SF4. The disappearance of this line indicates a phase transition of Cs$_4$PbBr$_6$ around 3 GPa, in agreement with previous report[43].

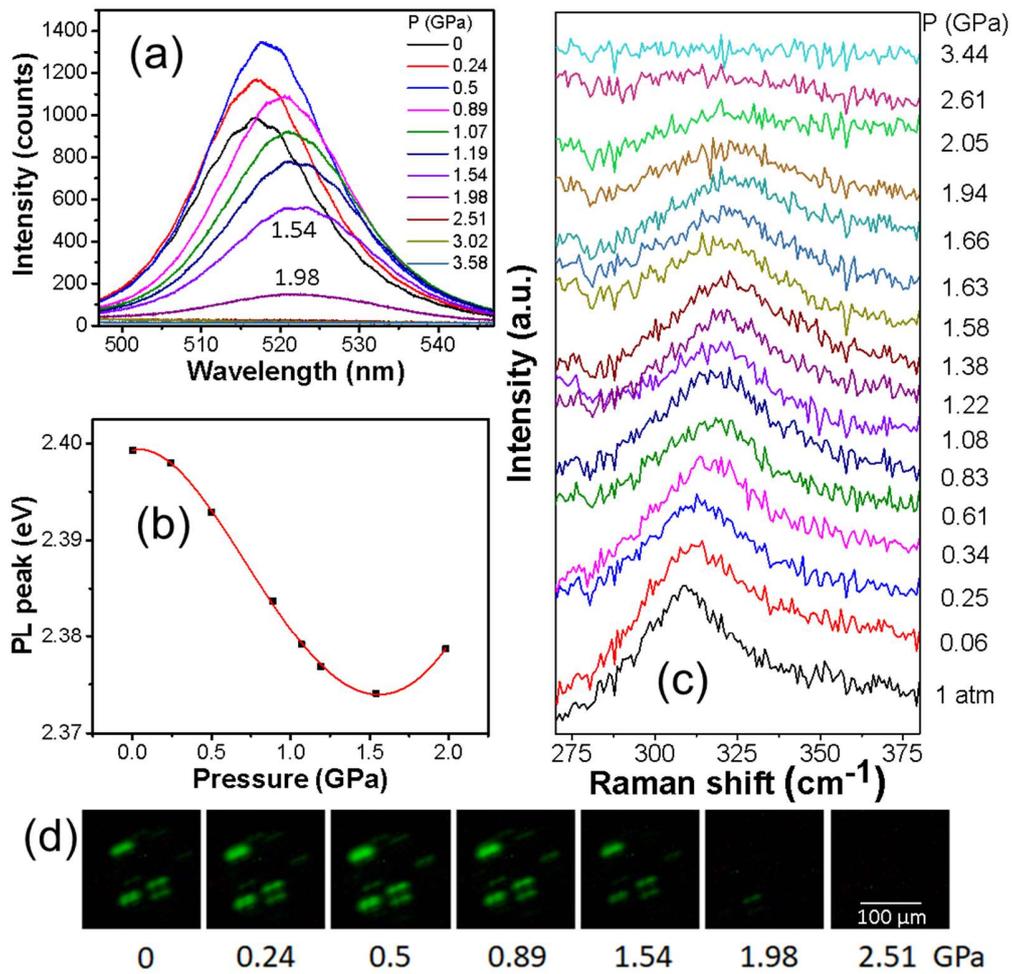

Fig. 4. (a-b) Evolution of PL spectrum and peak position of emissive Cs$_4$PbBr$_6$ crystals under hydrostatic pressure. (c) Zoomed-in view of evolution of Raman spectrum near 300 cm$^{-1}$ from Fig. 3d. (d) Optical images of green emission under increasing pressure.

Figure 4 shows the corresponding pressure evolution of green PL and 310 cm$^{-1}$ Raman band. The PL intensity increases initially and then decreases gradually until a quick disappearance at ~2.0 GPa. In the meantime, the PL peak has been shifting to longer wavelength until a turn to shorter wavelength at ~2.0 GPa. The 310 cm$^{-1}$ Raman band follows the same trend and disappears at around the same pressure of 2.0 GPa. These PL and Raman responses to pressure are nearly identical to that of CsPbBr$_3$ nanocrystals, as reported in the literature[49, 50] as well as observed in our own measurement (Figs. SF5-8). The strong correlation between PL and Raman, *i.e.*, the diminishing and disappearance of PL and Raman at the same pressure reveals that CsPbBr$_3$ nanocrystals are the sole source for the green PL.

We further argue that the observed pressure responses of PL and Raman cannot be explained by Br vacancies. The effect of hydrostatic pressure on point-defect states or luminescent centers has been investigated in many semiconductors[51, 52, 53, 54, 55, 56]. The consensus is that localized states such as Br vacancies are very stable under hydrostatic pressure, especially if they are already deep level states. Firstly, their optical activity will remain more or less the same and certainly will not disappear suddenly as long as its host material maintains the same structure. Secondly, the pressure induced energy shift should be less than the bandgap shift of the host material. The bandgap of Cs$_4$PbBr$_6$ decreases only by 10 meV under 1 GPa[43], but the observed PL shift is 20 meV, that is, twice the bandgap shift. As revealed both by Raman and bandgap evolution, Cs$_4$PbBr$_6$ remains stable until the phase transition at 3 GPa, but the PL already disappears at 2 GPa, way before the phase transition. Based on these two reasons, we can safely rule out Br vacancies as the source for green PL in Cs$_4$PbBr$_6$.

Having concluded that CsPbBr$_3$ nano-inclusions are the sole PL source, we comment on their size and try to understand some minor but confusing observations. Unlike CsPbBr$_3$ colloids or free-standing nanocrystals in Fig. SF5 and Ref. [[49, 50]], we always observe a PL enhancement under initial small pressure as shown in Fig. 4a. We believe that this is due to different dielectric environment or local strain experienced by the embedded CsPbBr$_3$ inclusions. Due to the same reason, the PL of CsPbBr$_3$ nano-inclusions exhibits a red shift compared to that of colloids[22, 23, 26, 57, 58]. Because of shorter PL peak wavelength, we believe that CsPbBr$_3$ nanocrystals in our case are smaller than reported nano-inclusions, which are less than 5 nm[33]. We also stress again that previous TEM evidence for pure emissive Cs$_4$PbBr$_6$ is not solidly justified, because it lacks a report of PL from the same nanocrystals before and after TEM imaging[7]. The observation of decreasing PL with increased $CsPbBr_3$ nanocrystal concentration in $Cs_4PbBr_6$ is due to the reduced PL quantum yield for larger size and lower quality $CsPbBr_3$[7].

One challenge for material sciences is that samples from different growers could be different even though they have followed the same recipes, *i.e.*, our emissive $Cs_4PbBr_6$ might not be exactly identical to those reported. However, our combined confocal Raman-PL and pressure Raman-PL techniques are generalizable. As demonstrated, they are sensitive and non-invasive, and can be easily adopted by any scientists with the proper equipment. In fact, in most cases, pressure Raman-PL with a DAC is not necessary because both Raman and PL can be quantitative if they are carefully calibrated against a reference sample such as $CsPbBr_3$ nanocrystals. A quick Raman can tell whether Br vacancies are solely responsible for the green PL, and a careful analysis of Raman and PL can tell the contribution of $CsPbBr_3$ nanocrystals to the total PL emission. A PL source other than $CsPbBr_3$ nanocrystals can be introduced only when the green PL cannot be accounted for by $CsPbBr_3$ nanocrystals with a near unity PL quantum yield.

**Conclusions**

We have identified $CsPbBr_3$ nanocrystal inclusions in $Cs_4PbBr_6$ as dominant green PL source in this compound. The complete Raman spectra of pristine $Cs_4PbBr_6$ are presented and the observed spectral lines are assigned to certain phonons in accordance with DFT lattice dynamics. We reveal the Raman signatures of emissive $Cs_4PbBr_6$. These are the doublet at 28-30 $cm^{-1}$ at low temperature and the Raman band at 310 $cm^{-1}$ at RT, shown to stem from $CsPbBr_3$ nanocrystal inclusions. No indication of Br vacancies related PL was found. The problem we faced in our attempt to find out whether intrinsic point defects can produce green PL in $Cs_4PbBr_6$ is that all tested emissive samples showed presence of $CsPbBr_3$. For future quests in searching for alternative origins of green PL in $Cs_4PbBr_6$ we suggest a mandatory Raman test. The resolution of this long-lasting controversy paves the way for device applications of low dimensional perovskites, and our comprehensive optical technique integrating structure-property with dynamic response can be applied to other materials to understand their luminescence centers.

**References:**

1. Zhao YX, Zhu K. Organic-inorganic hybrid lead halide perovskites for optoelectronic and electronic applications. *Chemical Society Reviews* **45**, 655-689 (2016).
2. Huang J, Yuan Y, Shao Y, Yan Y. Understanding the physical properties of hybrid perovskites for photovoltaic applications. *Nat Rev Mater* **2**, 17042 (2017).
3. Xiao JW, *et al.* The Emergence of the Mixed Perovskites and Their Applications as Solar Cells. *Adv Energy Mater* **7**, 1700491 (2017).
4. Zhu HM, *et al.* Lead halide perovskite nanowire lasers with low lasing thresholds and high quality factors. *Nat Mater* **14**, 636 (2015).
5. Saidaminov MI, *et al.* Pure $Cs_4PbBr_6$: Highly Luminescent Zero Dimensional Perovskite Solids. *ACS Energy Lett* **1**, 840-845 (2016).
6. Seth S, Samanta A. Photoluminescence of Zero-Dimensional Perovskites and Perovskite-Related Materials. *J Phys Chem Lett* **9**, 176-183 (2018).
7. Yin J, *et al.* Point Defects and Green Emission in Zero-Dimensional Perovskites. *J Phys Chem Lett* **9**, 5490-5495 (2018).
8. Akkerman QA, Abdelhady AL, Manna L. Zero-Dimensional Cesium Lead Halides: History, Properties, and Challenges. *J Phys Chem Lett* **9**, 2326-2337 (2018).
9. Han D, *et al.* Unraveling luminescence mechanisms in zero-dimensional halide perovskites. *Journal of Materials Chemistry C* **6**, 6398-6405 (2018).
10. Zhang XL, *et al.* All-Inorganic Perovskite Nanocrystals for High-Efficiency Light Emitting Diodes: Dual-Phase $CsPbBr_3$-$CsPb_2Br_5$ Composites. *Adv Funct Mater* **26**, 4595-4600 (2016).
11. Turedi B, *et al.* Water-Induced Dimensionality Reduction in Metal-Halide Perovskites. *J Phys Chem C* **122**, 14128-14134 (2018).
12. Li GP, *et al.* Shape and phase evolution from $CsPbBr_3$ perovskite nanocubes to tetragonal $CsPb_2Br_5$ nanosheets with an indirect bandgap. *Chem Commun* **52**, 11296-11299 (2016).
13. Wang Y, *et al.* Bright Luminescent Surface States on the Edges of Wide-bandgap Two-dimensional Lead Halide Perovskite. *arXiv preprint arXiv:180311490*, (2018).
14. Wang KH, Wu L, Li L, Yao HB, Qian HS, Yu SH. Large-Scale Synthesis of Highly Luminescent Perovskite-Related $CsPb_2Br_5$ Nanoplatelets and Their Fast Anion Exchange. *Angew Chem-Int Edit* **55**, 8328-8332 (2016).
15. Bao J, Hadjiev VG. Origin of Luminescent Centers and Edge States in Low-Dimensional Lead Halide Perovskites: Controversies, Challenges and Instructive Approaches. *Nano-Micro Letters* **In press**, DOI: 10.1007/s40820-40019-40254-40824 (2019).
16. Liu ZK, *et al.* Ligand Mediated Transformation of Cesium Lead Bromide Perovskite Nanocrystals to Lead Depleted $Cs_4PbBr_6$ Nanocrystals. *J Am Chem Soc* **139**, 5309-5312 (2017).
17. de Weerd C, Lin JH, Gomez L, Fujiwara Y, Suenaga K, Gregorkiewicz T. Hybridization of Single Nanocrystals of $Cs_4PbBr_6$ and $CsPbBr_3$. *J Phys Chem C* **121**, 19490-19496 (2017).
18. Yang L, Li DM, Wang C, Yao W, Wang H, Huang KX. Room-temperature synthesis of pure perovskite-related $Cs_4PbBr_6$ nanocrystals and their ligand-mediated evolution into highly luminescent $CsPbBr_3$ nanosheets. *J Nanopart Res* **19**, 258 (2017).
19. Iyikanat F, Sari E, Sahin H. Thinning $CsPb_2Br_5$ perovskite down to monolayers: Cs-dependent stability. *Phys Rev B* **96**, 155442 (2017).

20. Wang WK, Wang DF, Fang F, Wang S, Xu GH, Zhang TJ. $CsPbBr_3/Cs_4PbBr_6$ Nanocomposites: Formation Mechanism, Large-scale and Green Synthesis, and Application in White Light-Emitting Diodes. *Cryst Growth Des* **18**, 6133-6141 (2018).
21. Quan LN, *et al.* Highly Emissive Green Perovskite Nanocrystals in a Solid State Crystalline Matrix. *Adv Mater* **29**, 1605945 (2017).
22. Xu JW, *et al.* Imbedded Nanocrystals of $CsPbBr_3$ in $Cs_4PbBr_6$: Kinetics, Enhanced Oscillator Strength, and Application in Light-Emitting. *Adv Mater* **29**, 1703703 (2017).
23. Chen XM, *et al.* Centimeter-Sized $Cs_4PbBr_6$ Crystals with Embedded $CsPbBr_3$ Nanocrystals Showing Superior Photoluminescence: Nonstoichiometry Induced Transformation and Light-Emitting Applications. *Adv Funct Mater* **28**, 1706567 (2018).
24. Li YX, Huang H, Xiong Y, Kershaw SV, Rogach AL. Reversible transformation between $CsPbBr_3$ and $Cs_4PbBr_6$ nanocrystals. *Crystengcomm* **20**, 4900-4904 (2018).
25. Cho J, Banerjee S. Ligand-Directed Stabilization of Ternary Phases: Synthetic Control of Structural Dimensionality in Solution-Grown Cesium Lead Bromide Nanocrystals. *Chem Mat* **30**, 6144-6155 (2018).
26. Riesen N, Lockrey M, Badeke K, Riesen H. On the origins of the green luminescence in the "zero-dimensional perovskite" $Cs_4PbBr_6$: conclusive results from cathodoluminescence imaging *Nanoscale* **11**, 3925-3932 (2019).
27. Wang Y, *et al.* Solution-Grown $CsPbBr_3/Cs_4PbBr_6$ Perovskite Nanocomposites: Toward Temperature-Insensitive Optical Gain. *Small* **13**, 1701587 (2017).
28. Akkerman QA, *et al.* Nearly Monodisperse Insulator $Cs_4PbX_6$ (X = Cl, Br, I) Nanocrystals, Their Mixed Halide Compositions, and Their Transformation into $CsPbX_3$ Nanocrystals. *Nano Lett* **17**, 1924-1930 (2017).
29. Zhang ZJ, *et al.* Aqueous Solution Growth of Millimeter-Sized Nongreen-Luminescent Wide Bandgap $Cs_4PbBr_6$ Bulk Crystal. *Cryst Growth Des* **18**, 6393-6398 (2018).
30. De Bastiani M, *et al.* Inside Perovskites: Quantum Luminescence from Bulk $Cs_4PbBr_6$ Single Crystals. *Chem Mat* **29**, 7108-7113 (2017).
31. Yin J, *et al.* Intrinsic Lead Ion Emissions in Zero-Dimensional $Cs_4PbBr_6$ Nanocrystals. *ACS Energy Lett* **2**, 2805-2811 (2017).
32. Cha JH, *et al.* Photoresponse of $CsPbBr_3$ and $Cs_4PbBr_6$ Perovskite Single Crystals. *J Phys Chem Lett* **8**, 565-570 (2017).
33. Chen DQ, Wan ZY, Chen X, Yuan YJ, Zhong JS. Large-scale room-temperature synthesis and optical properties of perovskite-related $Cs_4PbBr_6$ fluorophores. *Journal of Materials Chemistry C* **4**, 10646-10653 (2016).
34. Bao Z, Wang HC, Jiang ZF, Chung RJ, Liu RS. Continuous Synthesis of Highly Stable $Cs_4PbBr_6$ Perovskite Microcrystals by a Microfluidic System and Their Application in White-Light-Emitting Diodes. *Inorg Chem* **57**, 13071-13074 (2018).
35. Seth S, Samanta A. Fluorescent Phase-Pure Zero-Dimensional Perovskite-Related $Cs_4PbBr_6$ Microdisks: Synthesis and Single-Particle Imaging Study. *J Phys Chem Lett* **8**, 4461-4467 (2017).
36. Chen X, Chen DQ, Li JN, Fang GL, Sheng HC, Zhong JS. Tunable $CsPbBr_3/Cs_4PbBr_6$ phase transformation and their optical spectroscopic properties. *Dalton Trans* **47**, 5670-5678 (2018).
37. Lv JF, Fang LL, Shen JQ. Synthesis of highly luminescent $CsPb_2Br_5$ nanoplatelets and their application for light-emitting diodes. *Mater Lett* **211**, 199-202 (2018).

38. Ruan LF, Lin J, Shen W, Deng ZT. Ligand-mediated synthesis of compositionally related cesium lead halide CsPb$_2$X$_5$ nanowires with improved stability. *Nanoscale* **10**, 7658-7665 (2018).
39. Zhang Y, *et al.* Ligand-Free Nanocrystals of Highly Emissive Cs$_4$PbBr$_6$ Perovskite. *J Phys Chem C* **122**, 6493-6498 (2018).
40. Stoumpos CC, *et al.* Crystal Growth of the Perovskite Semiconductor CsPbBr$_3$: A New Material for High-Energy Radiation Detection. *Cryst Growth Des* **13**, 2722-2727 (2013).
41. Dong QF, *et al.* Electron-hole diffusion lengths > 175 mm in solution-grown CH$_3$NH$_3$PbI$_3$ single crystals. *Science* **347**, 967-970 (2015).
42. Bao JM, *et al.* Optical properties of rotationally twinned InP nanowire heterostructures. *Nano Lett* **8**, 836-841 (2008).
43. Ma ZW, *et al.* Pressure-induced emission of cesium lead halide perovskite nanocrystals. *Nat Commun* **9**, 4506 (2018).
44. Cola M, Massarot.V, Riccardi R, Sinistri C. Binary systems formed by lead bromide with (Li, Na, K, Rb, Cs and Tl)Br - DTA and diffractometric study. *Zeitschrift Fur Naturforschung Part a-Astrophysik Physik Und Physikalische Chemie* **A 26**, 1328 (1971).
45. Rousseau DL, Bauman RP, Porto SPS. Normal mode determination in crystals. *J Raman Spectrosc* **10**, 253-290 (1981).
46. Hayes W, Loudon R. *Scattering of Light by Crystals*. New York : Wiley (1978).
47. Yaffe O, *et al.* Local Polar Fluctuations in Lead Halide Perovskite Crystals. *Phys Rev Lett* **118**, 136001 (2017).
48. Hadjiev VG, *et al.* Phonon fingerprints of CsPb$_2$Br$_5$. *J Phys-Condes Matter* **30**, 405703 (2018).
49. Xiao GJ, *et al.* Pressure Effects on Structure and Optical Properties in Cesium Lead Bromide Perovskite Nanocrystals. *J Am Chem Soc* **139**, 10087-10094 (2017).
50. Zhang L, Zeng QX, Wang K. Pressure-Induced Structural and Optical Properties of Inorganic Halide Perovskite CsPbBr$_3$. *J Phys Chem Lett* **8**, 3752-3758 (2017).
51. Haller EE, Hsu L, Wolk JA. Far infrared spectroscopy of semiconductors at large hydrostatic pressures. *Phys Status Solidi B-Basic Res* **198**, 153-165 (1996).
52. Xue YZ, *et al.* Anomalous Pressure Characteristics of Defects in Hexagonal Boron Nitride Flakes. *ACS Nano* **12**, 7127-7133 (2018).
53. Sha XJ, *et al.* Ab initio study of native point defects in ZnO under pressure. *Solid State Commun* **201**, 130-134 (2015).
54. Millot M, Geballe ZM, Yu KM, Walukiewicz W, Jeanloz R. Red-green luminescence in indium gallium nitride alloys investigated by high pressure optical spectroscopy. *Appl Phys Lett* **100**, 162103 (2012).
55. Strikha MV, Vasko FT. Hydrostatic-pressure effect on point-defect electronic states in narrow-gap and gapless semiconductors. *Phys Status Solidi B-Basic Res* **181**, 181-188 (1994).
56. Zhao YC, Barvosa-Carter W, Theiss SD, Mitha S, Aziz MJ, Schiferl D. Pressure measurement at high temperature using ten Sm : YAG fluorescence peaks. *J Appl Phys* **84**, 4049-4059 (1998).
57. Protesescu L, *et al.* Nanocrystals of Cesium Lead Halide Perovskites (CsPbX$_3$, X = Cl, Br, and I): Novel Optoelectronic Materials Showing Bright Emission with Wide Color Gamut. *Nano Lett* **15**, 3692-3696 (2015).

58. Brennan MC, *et al.* Origin of the Size-Dependent Stokes Shift in CsPbBr$_3$ Perovskite Nanocrystals. *J Am Chem Soc* **139**, 12201-12208 (2017).